\documentclass[aps,prl,twocolumn,showpacs,superscriptaddress]{revtex4-1}

\usepackage[english]{babel}
\usepackage{graphicx}
\usepackage{dcolumn}
\usepackage{amsmath}
\usepackage{epsfig}
\usepackage{bm}
\usepackage{amssymb}
\usepackage{float}
\usepackage{natbib}
\usepackage{color}
\usepackage{soul}
\usepackage{hyperref}
\hypersetup{colorlinks=true,linkcolor=blue,citecolor=blue,urlcolor=black}

\usepackage{units}

\usepackage{hyperref}
\hypersetup{%
colorlinks,
linkcolor={black},
citecolor={black},
urlcolor={black}}

\newcommand{\ket}[1]{| #1 \rangle}

\begin{document}

\title{Preparation of Long-Lived, Non-Autoionizing Circular Rydberg States of Strontium}

\author{R. C. Teixeira}
\affiliation{Laboratoire Kastler Brossel, Coll\`ege de France,
  CNRS, ENS-Universit\'e PSL,
  Sorbonne Universit\'e,  \\11, place Marcelin Berthelot, 75005 Paris, France}
  
  \author{A. Larrouy}
\affiliation{Laboratoire Kastler Brossel, Coll\`ege de France,
  CNRS, ENS-Universit\'e PSL,
  Sorbonne Universit\'e,  \\11, place Marcelin Berthelot, 75005 Paris, France}
  
  \author{A. Muni}
\affiliation{Laboratoire Kastler Brossel, Coll\`ege de France,
  CNRS, ENS-Universit\'e PSL,
  Sorbonne Universit\'e,  \\11, place Marcelin Berthelot, 75005 Paris, France}
  
  \author{L. Lachaud}
\affiliation{Laboratoire Kastler Brossel, Coll\`ege de France,
  CNRS, ENS-Universit\'e PSL,
  Sorbonne Universit\'e,  \\11, place Marcelin Berthelot, 75005 Paris, France}

  \author{J.-M. Raimond}
\affiliation{Laboratoire Kastler Brossel, Coll\`ege de France,
  CNRS, ENS-Universit\'e PSL,
  Sorbonne Universit\'e,  \\11, place Marcelin Berthelot, 75005 Paris, France}
  
  \author{S. Gleyzes}
  \affiliation{Laboratoire Kastler Brossel, Coll\`ege de France,
  CNRS, ENS-Universit\'e PSL,
  Sorbonne Universit\'e,  \\11, place Marcelin Berthelot, 75005 Paris, France}

\author{M. Brune}
\affiliation{Laboratoire Kastler Brossel, Coll\`ege de France,
  CNRS, ENS-Universit\'e PSL,
  Sorbonne Universit\'e,  \\11, place Marcelin Berthelot, 75005 Paris, France}

\hyphenation{Ryd-berg sen-sing ma-ni-fold}

\date{\today}

\begin{abstract}
Alkaline earth Rydberg atoms are very promising tools for quantum technologies. Their highly excited outer electron provides them with the remarkable properties of Rydberg atoms and, notably, with a huge coupling to external fields or to other Rydberg atoms while the ionic core retains an optically active electron. However, low angular-momentum Rydberg states suffer almost immediate autoionization when the core is excited. Here, we demonstrate that strontium circular Rydberg atoms with a core excited in a $4D$ metastable level are impervious to autoionization over more than a few millisecond time scale. This makes it possible to trap and laser-cool Rydberg atoms.  Moreover, we observe singlet to triplet transitions due to the core optical manipulations, opening the way to a quantum microwave to optical interface.
\end{abstract}

\maketitle

Rydberg atoms are a promising platform for quantum information \cite{saffman_quantum_2010}, quantum metrology \cite{fan_atom_2015, facon_sensitive_2016} or quantum simulation \cite{Labuhn_tunable_2016, bernien_probing_2017}. For a long time, most Rydberg atom experiments have been using alkali atoms. However, alkaline-earth Rydberg atoms are now the focus of an intense experimental activity \cite{dunning_recent_2016}. In the context of alkaline-earth-based optical atomic clocks, dressing with or excitation to Rydberg states open promising perspectives for noise reduction by the preparation of non-classical states of an atomic ensemble \cite{gil_spin_2014,Mukherjee_many-body_2011}.
The spectroscopy of  Rydberg states also provides accurate methods to estimate the systematic shift of these clocks induced by the blackbody radiation or by residual electric fields \cite{Ovsiannikov_rydberg_2011,bowden_rydberg_2017,cantat-moltrecht_long-lived_2020}.
More importantly, once in the Rydberg state, alkaline-earth atoms feature an optically active ionic core that makes it possible to image \cite{Lochead_number-resolved_2013,Mcquillen_imaging_2013} or trap \cite{bounds_Rydberg-Dressed_2018,wilson_trapped_2019} them, opening interesting perspectives for quantum simulation \cite{cooper_alkaline-earth_2018, norcia_microscopic_2018}. 

Most alkaline-earth experiments so far access low-angular-momentum Rydberg states~\cite{millen_spectroscopy_2011,ye_production_2013,ye_efficient_2014}, for which an excitation of the ionic core electron quickly leads to autoionization~\cite{millen_two-electron_2010,fields_autoionization_2018}. 
In order to overcome this limit, it is mandatory to increase the angular momentum $\ell$ of the Rydberg electron  \cite{Mcquillen_imaging_2013}. 
Up to now, 
moderately large-$\ell$ states 
have been produced using microwave transfer \cite{Niyaz_microwave_2019} or Stark switching methods \cite{cooke_doubly_1978,jones_autoionization_1988,pruvost_high_1991,wehrli_autoionization_2019}. These early studies have evidenced the reduction of the autoionization rate with $\ell$ but extending their techniques to $\ell>10$ is challenging \cite{lehec_isolated_2020}. 

The circular states of alkaline-earth atoms are expected to offer the best protection against autoionization. Circular Rydberg levels \cite{hulet_rydberg_1983,signoles_coherent_2017} have the maximum allowed magnetic quantum number, $m=\ell=n-1$. They have remarkable properties, in particular a very long intrinsic lifetime (30 ms for $n=50$). They were used extensively to manipulate the microwave fields of superconducting cavities \cite{raimond_manipulating_2001}. They have recently attracted a broader interest, from cold atom physics \cite{kleinbach_ionic_2018} to quantum metrology \cite{ramos_measuring_2017,facon_sensitive_2016} or quantum simulation \cite{nguyen_towards_2018}. Finally, their wave-function is localized at a large distance from the nucleus. Hence, the outer electron of an alkaline-earth circular state has a very small overlap with the ionic core and the autoionization rate is expected to be extremely low. An early study of the $n=21$ circular state of Barium had already exhibited reduced autoionization for a core in one of its metastable levels~\cite{roussel_observation_1990}. Their low autoionization rate, combined with their unique properties, explains the interest for circular states of alkaline earth in recent proposals~\cite{wilson_trapped_2019,meinert_indium_2020}.

In this Letter, we report an important experimental step in the control of alkaline earth Rydberg atoms, with the preparation of the $n=51$ singlet circular state of strontium. We observe transitions to the triplet states resulting from a selective manipulation of the ionic core electron spin. We observe that autoionization is negligible when the ionic core electron is promoted to any of the $4D$ metastable states. These results opens bright perspectives for the optical manipulation of circular Rydberg states.

\begin{figure}
  \centering
  \includegraphics[width=.8\linewidth]{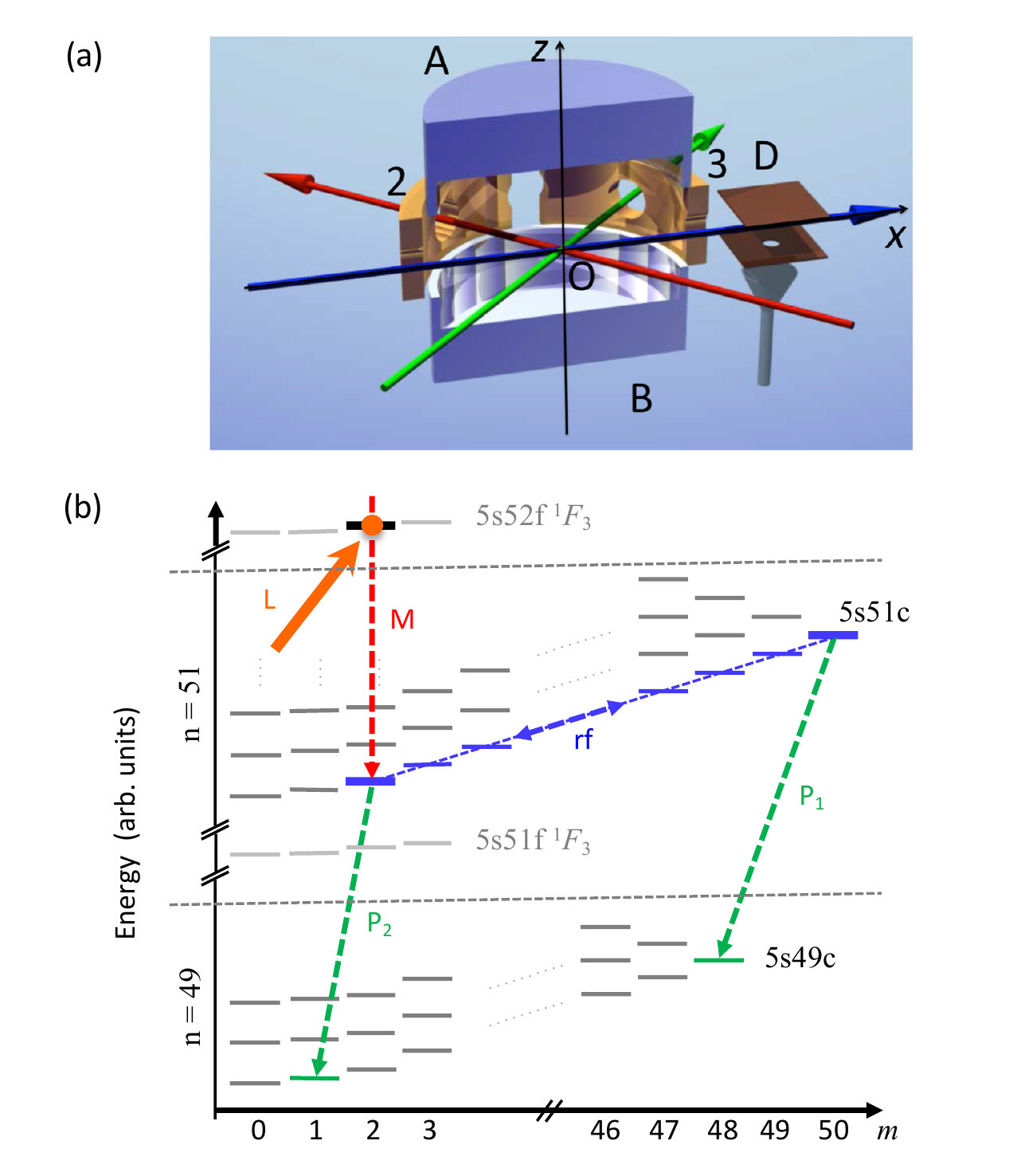}
  \caption{Circular states preparation and probe.  (a) Experimental setup. A thermal atomic beam (blue arrow) intersects the laser beams (767 nm and 896 nm in green, 461 and 422 nm in red) at the center of an electrode structure.  The electrodes $A$ and $B$ create the vertical field along the $Oz$ axis. Four ring electrodes 1, 2, 3 and 4 (1 and 4 not shown) create a $\sigma^+$-polarized rf field and a static electric field along the direction of the 461nm laser beam. The atoms are finally detected in the ionization detector $\cal{D}$.
  (b) Stark energy levels of the relevant Rydberg manifolds. The atoms are laser-excited into the $m=2$ sublevel of the $5s52f\ ^1F_3 $ state. They are transferred into the lowest $m=2$ state of the $n=51$ manifold by a preparation mw pulse M. A $\sigma^+$-polarized rf  circularization pulse (dashed blue line) resonant with the transitions between the Stark levels couples the states of the lower diagonal of the $n=51$ manifold (blue levels) and transfers the atom into the circular state $5s51c$. We measure the preparation fidelity by selectively transferring $5s51c$ to $5s49c$ using  a probe mw $\pi$-pulse P$_1$. For measuring the populations of the circular singlet/triplet states, we apply sucessively a decircularization rf pulse that transfers the atom back toward $m=2$  and a probe mw $\pi$-pulse P$_2$ toward the lowest $m=1$ state of the 49 manifold. This pulse resolves singlet and triplet transitions.}
  \label{fig:inst_eigenvalsc}
\end{figure}

The strontium circular Rydberg states are produced from a thermal atomic beam inside an electrode structure controlling the static electric field at the atomic position [Fig. 1(a)]. The experiment is cooled down to 4.2~K by a wet $^4$He cryostat. The atoms are promoted from the $5s^2\ ^1S_0$ singlet ground state to the $5s52f\ ^1F_3 $ Rydberg state by a stepwise resonant three-photon excitation (461 nm, 767 nm and 896 nm) that uses $5s5p\ ^1P_1$ and $5s4d\ ^1D_2$ as intermediate states \cite{SI}. The laser beams are sent at a $45^\circ$ angle with respect to the atomic beam, allowing us to select the atomic velocity $v=418\pm 9$~m/s using Doppler effect. The 767 nm and 896 nm beams are collinear and sent perpendicular to the 461 nm beam. The 896 nm laser is pulsed for 1~$\mu$s at the beginning of the sequence, the other lasers are cw. During laser excitation, a small static electric field ($\sim 0.6$ V/cm) applied by the ring electrodes circling the structure [Fig.~\ref{fig:inst_eigenvalsc}(a)] sets a quantization axis aligned with the direction of the 461 nm laser beam and lifts the degeneracy between the sublevels of the $52f$ states with different $|m|$ values. We set the 896 nm laser on resonance with the transition toward $|m| = 2$. The $\sigma^+$ polarization of the 461 nm laser selects the $m=2$ level. Once the atoms are excited into Rydberg sates, a 422 nm core excitation laser pulse (direction parallel to the 461 nm laser beam) can be used for exciting the second valence electron at a wavelength close to the $5S_{1/2}$ and  $5P_{1/2}$ resonance of Sr$^+$.

For alkali atoms \cite{signoles_coherent_2017}, the  transfer from the laser-accessible state to the circular state involves a series of degenerate radio-frequency (rf) transitions between the Stark levels. For strontium, Fig. 1(b) presents the relevant Rydberg levels in a static electric field. Due to its large quantum defect, the laser-excited $5s52f\ ^1F_3 $ level is far from the hydrogenic manifold for fields on the order of $F \sim 1$ V/cm. We thus transfer  the atom from this level into the lowest $m=2$ state of the $n=51$ Stark manifold with a first microwave (mw) $\pi$-pulse. We then rotate adiabatically the electric field and align it along the vertical axis $Oz$. The ring electrodes circling the set-up finally generate a $\sigma^+$-polarized rf  `circularization' pulse, resonant with the transitions between the lowest levels in the Stark ladder [blue states  and arrows on Fig.~1(b)].  This pulse transfers the population of the $m=2$ level into the circular state. The process starts from a laser excited singlet state. Since the mw and rf fields do not change the electron spin, the atoms are finally prepared in the singlet state $5s51c, S=0$.

Fig.~2(a) shows a generic timing of the experiment. Fig.~2(b) shows the ionization signals measured with the state-selective field-ionization detector $\cal{D}$ at different steps of the preparation. Low-$m$ levels prepared by the preparation microwave pulse M ionize in a field of about 50 V/cm (green line). When we apply the circularization rf pulse, the atoms ionize in a much larger field, close to the expected ionization threshold of the circular state (black line). The residual ionization peak at low field is mainly due to atoms left in the $5s52f\ ^1F_3 $ state due to the imperfection of the mw pulse M. This shows that most of the population in the $n=51$ manifold has been transferred to high-angular momentum states.  Because of the anharmonicity of the Stark level structure for low-$m$ states, the resonant rf pulse cannot prepare the circular state with unit efficiency \cite{signoles_coherent_2017}. In order to measure the number of atoms in the circular state, we apply a mw probe pulse P$_1$ (red line). In the applied $F= 1.4$~V/cm electric field, it is resonant with the two-photon transition between the $5s51c$ and $5s49c$ circular states but out-of-resonance for atoms  in non-circular states {[see Fig. 1(b)]. The large transfer rate to the $n=49$ manifold  shows that at most 15 \% of the atoms are prepared in non-circular high-$\ell$ states.

\begin{figure}
  \centering
  \includegraphics[width=\linewidth]{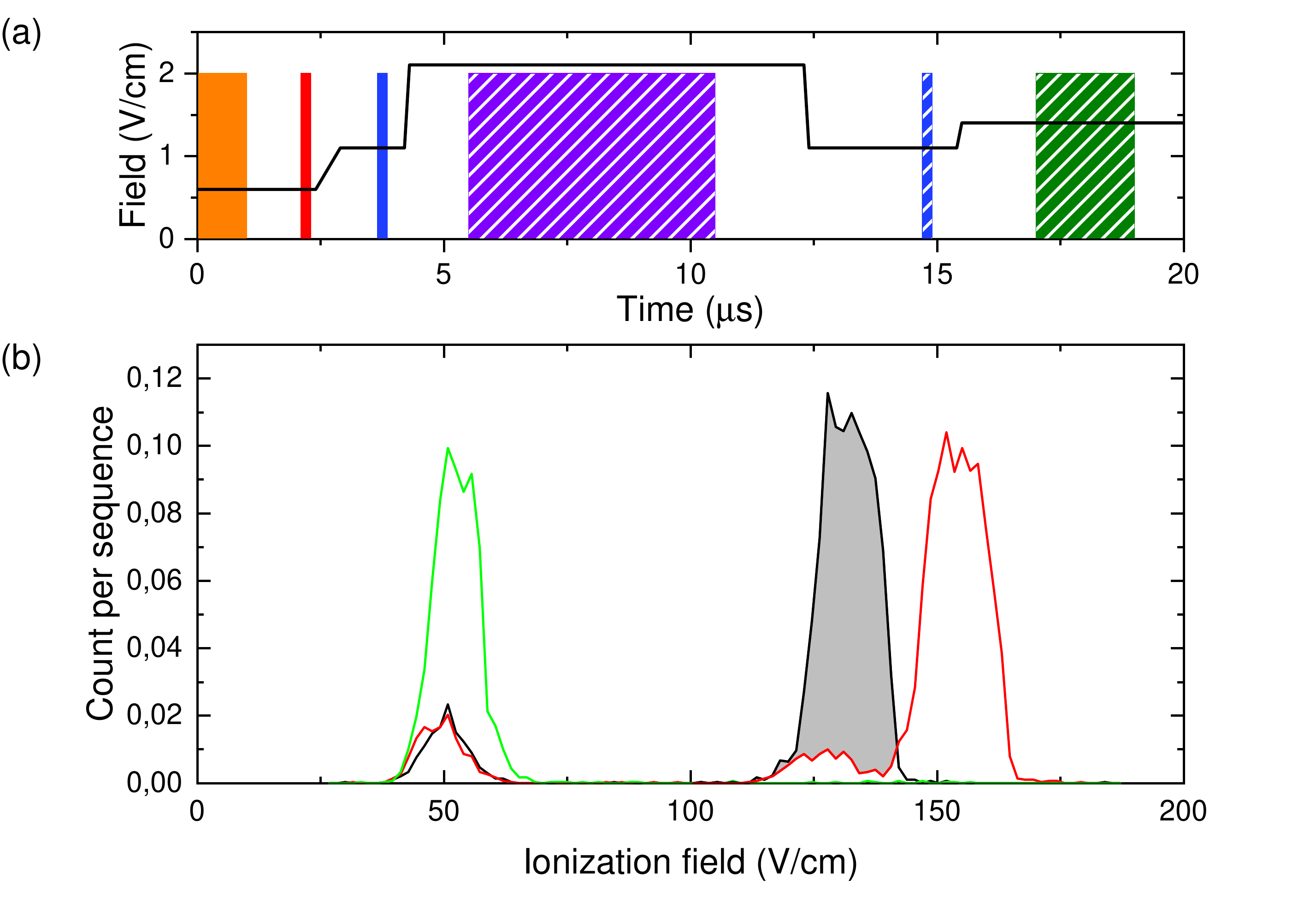}\\
  \caption{Circular state preparation.  (a) Experimental timing. Orange, red, blue, violet and green areas correspond to laser excitation, mw preparation, $\sigma^+$-polarized rf pulse, 422nm  laser and mw probe pulses respectively. Events with hatched area are not applied for the data presented in figure 2(b). (b) Ionization signals. Number of detected atoms as a function of the ionization voltage after microwave preparation pulse M pulse (green), after the circularization rf pulse (black), after the rf pulse and the probe mw pulse P1 (red). Atoms detected close to the ionization threshold of the $51c$ level in the red curve correspond to elliptical levels not transferred to $49c$ by the microwave probe P$_1$. The difference between the black and red curves (shaded area) corresponds to the atoms in the $51c$ circular state.}
  \label{fig:dig-timing}
\end{figure}

We then check that the circular states are insensitive to autoionization when we optically excite the ionic core. The far-away Rydberg electron has a negligible influence on the ionic core. We thus expect the core level structure to be nearly identical that of the Sr$^+$ ion  \cite{fields_autoionization_2018}. The doubly excited states have thus the form $j_1,l_1,51c$, where $j_1,l_1$ refer to the total and orbital angular momenta of the ionic core electron. After the rf pulse, we excite the core with the 422 nm laser beam tuned close to the transition frequency between the $5S_{1/2}$ and  $5P_{1/2}$ levels of Sr$^+$. The beam has a gaussian waist of 430 $\mu$m and a maximum power $P$ of 16~mW. The interaction time is limited by the atomic transit across the beam.  When the laser is on resonance, the atoms excited to $5P_{1/2},51c$ can decay into the $4D_{3/2}, 51c$ metastable state. The $4D_{3/2}$ level of Sr$^+$ has a lifetime of 0.4 s. The circular state electron is not expected to change this lifetime by much \cite{Mcquillen_imaging_2013}. For a large enough $P$, atoms are expected to be optically pumped in this state and remain in it for the whole time of flight toward detector $\cal D$ (140 $\mu$s).

\begin{figure}
  \centering
  \includegraphics[width=\linewidth]{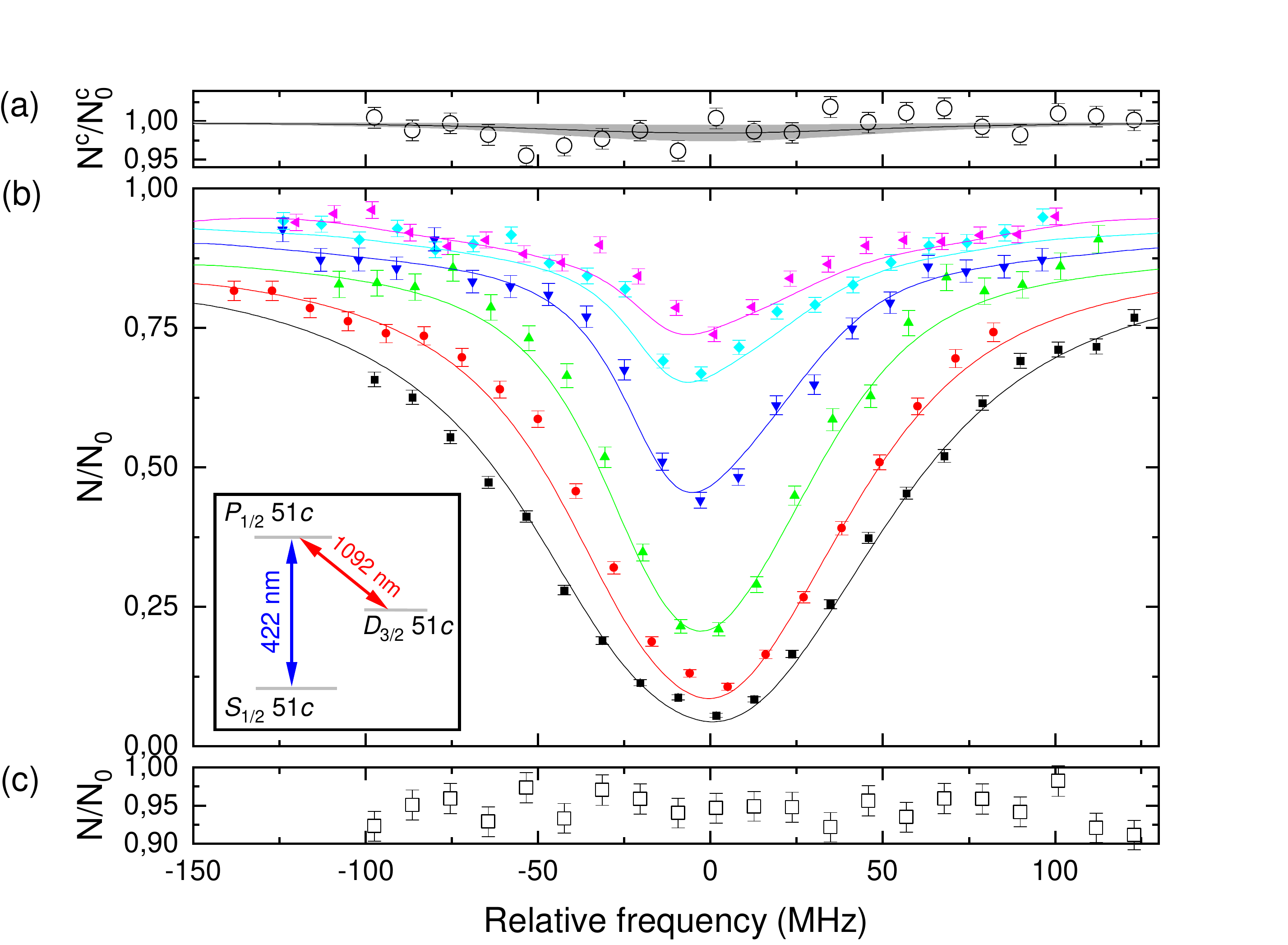}\\ 
     \includegraphics[width=\linewidth]{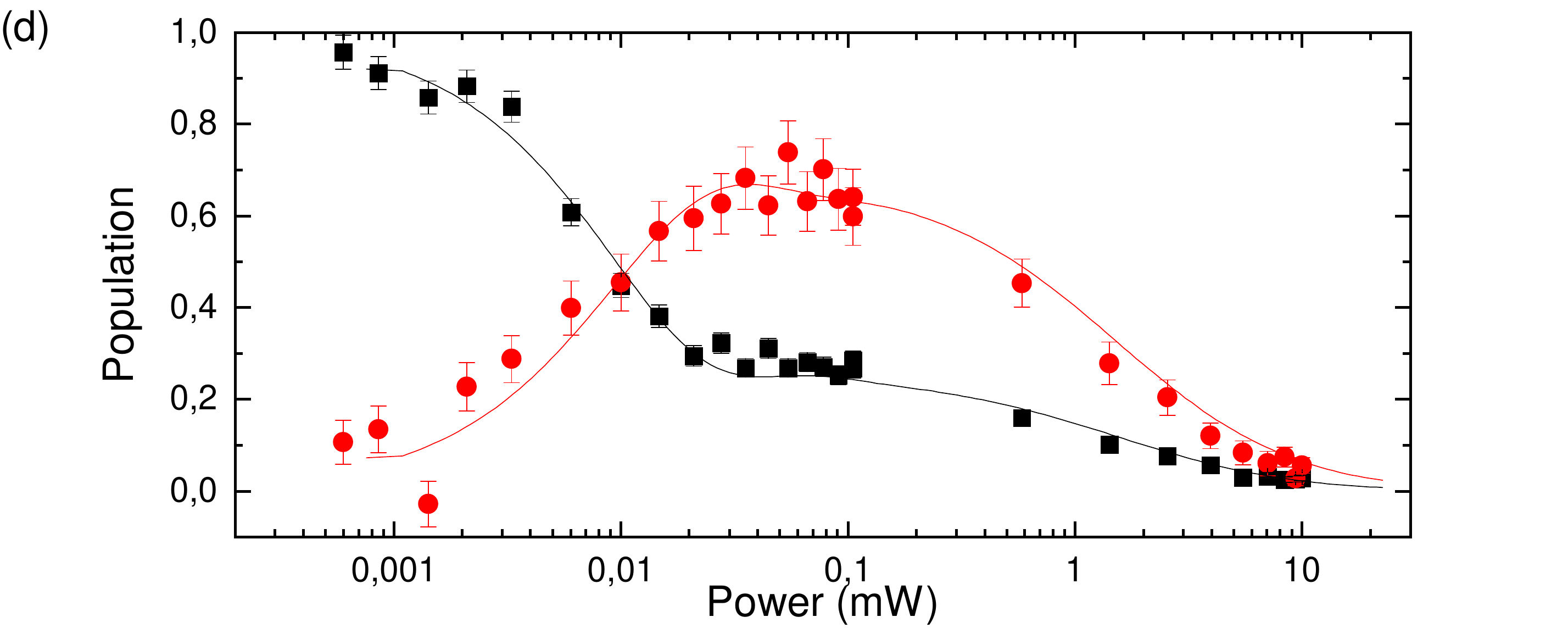}\\
  \caption{Interaction with the 422 nm core excitation laser. (a) Normalized number $N^c/N^c_0$ of atoms detected in the circular state 51c after applying the 422 nm laser pulse as a function of its relative frequency $\Delta\nu_\mathrm{l}$. The laser power is $P=$~16 mW. The normalization factor $N^c_0$ is the number of 51c atoms measured without this laser pulse. 
 The frequency reference is arbitrary. The points are experimental with statistical error bars. The shadow area correspond to the 95~\% confidence band of the fit of the survival probability (solid line, see text). (b) Normalized number $N/N_0$ of atoms detected in low-$m$ states after decircularization rf pulse as function of $\Delta\nu_\mathrm{l}$ for different laser powers (black: 16 mW, red: 9.5 mW, green: 4.0 mW, blue: 1.3 mW, cyan: 0.5 mW, magenta: 0.3 mW). 
 The inset shows the relevant levels. (c) Normalized number $N/N_0$ for $P=$~16 mW with 1092 nm repumper light. (d) Population of the $5S_{1/2}, 51c$ singlet ($p_S$, black) and triplet ($p_T$, red) states  as a function the 422 nm laser power $P$ when it is is set at resonance. (b-d) The points are experimental (with statistical error bars) and the lines are the predictions of the numerical model.}
  \label{fig:spectro}
\end{figure}

Fig. \ref{fig:spectro}(a) presents the population left in the circular state as a function of the relative frequency $\Delta\nu_\mathrm{l}$ of the 422 mn laser. It is nearly constant within the precision of the measurement. This shows that autoionization is negligible on a 140 $\mu$s timescale. 

We now indirectly probe the fraction of the atoms optically pumped in $4D_{3/2}, 51c$ by applying a `decircularization' rf pulse [hatched blue rf pulse on figure 2(a)] that transfers the circular state back to  low-$m$ states, which autoionize immediately when the ionic core is excited. We thus only count in $\cal D$ those atoms whose core remained unexcited ($5S_{1/2}$) after the laser pulse.  
Fig. \ref{fig:spectro}(b) presents the number $N$, of $n=51$ Rydberg atoms detected at the $m=2$ ionization threshold after the decircularization rf pulse, normalized by the number $N_0$, of atoms  detected without the 422 nm laser \cite{SI}. As expected, we observe a resonance dip, whose depth increases with the blue laser power $P$. As an additional check, we perform the same experiment with an additional 1092-nm ``repumper'' light, resonant with the $4D_{3/2}, 51c\rightarrow 5P_{1/2},51c$ transition [see level diagram in the inset of Fig.~\ref{fig:spectro}(b)]. It efficiently empties $4D_{3/2}, 51c$  \cite{SI}. We then observe a flat signal, close to 100 \% [Fig.~\ref{fig:spectro}(c)], confirming that the dip in  Fig. \ref{fig:spectro}(b) is due to optical pumping in $4D_{3/2}, 51c$. 

The data are in excellent agreement with numerical simulations computing the population of $5S_{1/2}, 51c$ before the decircularization rf pulse [solid lines in Fig.~\ref{fig:spectro}(b)]. They are based on a detailed model of the optical pumping induced by the 422 nm laser beam within the magnetic sublevels of the $S$ and $P$ core states. We include in these simulations the atomic velocity distribution and the spatial extension of the atomic sample. The only adjustable parameter is the proportionality factor between the 422 nm laser power $P$ and its intensity at the position of the atoms, which we fit to be $26 \%$  smaller than the expected value \cite{SI}.  Due to the spin-orbit coupling of the ionic core electron in the $5P_{1/2},51c$ state, scattering photons at 422 nm may flip the spin of the ionic core electron, and thus induce singlet to triplet transition for the atom. For high-$m$ Rydberg states, triplet and singlet states are degenerate, but for low-$m$ states, the exchange energy is no longer negligible. Hence, the transfer efficiency of the decircularization rf pulse is different for an atom in the singlet or triplet state. We take this effect into account by including in the model a detection efficiency of the $5S_{1/2}, 51c$ triplet states equal to 93\% of that of the singlet state.

The exchange energy for low-$m$ states makes it possible to use the Rydberg electron to probe the core electron spin dynamics. We apply after the decircularization pulse an engineered selective mw probe pulse [green rectangle on figure 2(a)] transferring either the singlet or triplet lowest state with $m=2$ in the $n=51$ manifold into the $n=49$ manifold . The number of atoms detected in the $n=49$ manifold after the probe pulse allows us to respectively reconstruct the population $p_S$ and $p_T$ of the singlet and triplet $5S_{1/2},51c$ states at the end of the optical pumping process \cite{SI}. Fig. 3(d) presents the evolution of these populations as a function of $P$ when the 422 nm laser is on resonance and the 1092 nm repumper is not applied. The solid lines present the predictions of the optical pumping model. They are in excellent agreement with the data.

We observe two regimes. Up to $P=0.1$ mW, the singlet state population decreases with $P$ down to $\sim 25\%$, while the population of the triplet state rises to $\sim 75\%$.  Indeed, the polarization of the 422 nm laser has been chosen to be nearly $\sigma^-$ (for the ionic core state, we consider a quantization axis along the direction of the 422 nm laser). Thus, the $5S_{1/2}, m_j= -1/2$ sublevel is almost a dark state \cite{berkeland_Destabilization_2002}, into which the core electron is optically pumped  after a few absorption-spontaneous emission cycles only. The final state of the optical pumping process is a 1:3 mixture of singlet and triplet states.

For higher powers, $P>0.1$ mW, the small $\sigma^+$ component of the 422 laser polarization (the ratio between the $\sigma^+$ and $\sigma^-$ intensities is measured to be 1:36) becomes large enough to have a sizable effect. The $5S_{1/2}, m_j= -1/2$ state is no longer a dark state. Absorption and spontaneous emission cycles go on until the atoms fall in the  $4D_{3/2}, 51c$  metastable state and are finally autoionized after the decircularization pulse. Accordingly, both singlet and triplet populations decrease with $P$. For the largest power, we estimate that 95 \% of the atoms are in the metastable core state before the decircularization pulse is applied.

The optical pumping model combined with the data of Fig. \ref{fig:spectro}(a) allows us to estimate 
a lower bound for the autoionization lifetime of $4D_{3/2}, 51c$. 
We fit on the signal of Fig. \ref{fig:spectro}(a) a dip proportional to the number of atoms in $4D_{3/2}, 51c$ provided by the model. 
We find a dip depth between 0.4 \% and 2.7 \%	with a 95 \% confidence, corresponding, for a time of flight of 140 $\mu s$ to an autoionization lifetime larger than 5 ms \cite{SI} .

\begin{figure}
  \centering
    \includegraphics[width=\linewidth]{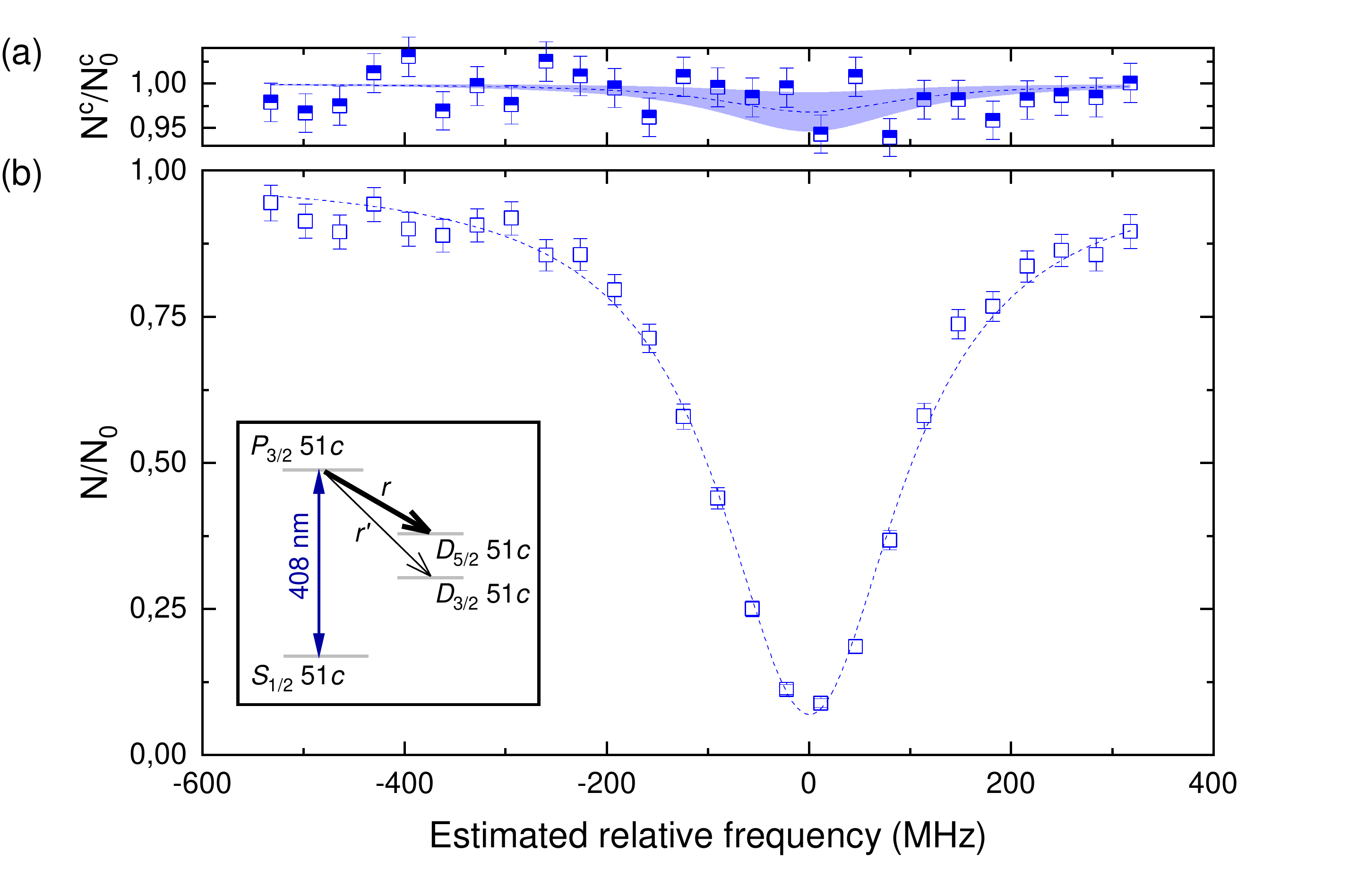}
  \caption{Survival probability $N^c/N^c_0$ (a) and $N/N_0$ (b) in the presence of 408 nm light as a function of the laser frequency. The points are experimental with statistical error bars. The dashed lines are lorentzian fits of both curves with the same width and center frequency. The shadow area in (a) corresponds to the 95~\% confidence band of the fit. The inset shows the relevant ionic core levels. }
  \label{fig:408}
\end{figure}

We finally briefly investigate the autoionization lifetime of $4D_{5/2},51c$. To that end, we perform a measurement similar to that of Fig. \ref{fig:spectro}(ab) with a 408 nm laser exciting the atom into the $5P_{3/2},51c$ level. The 408~nm laser is a free-running frequency-doubled Ti:Sa laser \cite{SI}. The laser has a power $P'$ on the order of a milliwatt, sufficient to optically pump the atom in the metastable state since the $5S_{1/2}-5P_{3/2}$ transition does not have a dark state for any polarization \cite{berkeland_Destabilization_2002}. 

 Fig.~\ref{fig:408} presents the circular state survival probability, as well as the number of atoms detected in low-$m$ states after the decircularization rf pulse, as a function of the 408 nm laser frequency. To eliminate the effect of the slow drift of this frequency \cite{SI}, the two curves are recorded simultaneously. As for Fig. \ref{fig:spectro}, we observe no significant variation on Fig.~\ref{fig:408}(a) and a deep dip in  Fig.~\ref{fig:408}(b) due to autoionization when we return to low-$m$ state after core excitation. Note that the $5P_{3/2}$ state decays into both $4D_{5/2}$ and $4D_{3/2}$ levels with rates $r$ and $r^\prime$ respectively [inset of Fig. 4(b)]. The branching ratio  $r/r^\prime=8.4$ measured in Sr$^+$  \cite{zhang_iterative_2016} indicates that the atoms are mostly pumped into  $4D_{5/2},51c$ \cite{SI}.

We get an estimate of the $4D_{5/2},51c$ autoionization lifetime, by fitting the data of  Fig.~\ref{fig:408}(b) with a lorentzian lineshape (dashed line). The dip depth corresponds to the fraction of the atoms optically pumped in the metastable states (within a factor close to 1, due to different detection efficiencies of the singlet and triplet states).  In order to estimate the survival probability of $4D_{5/2},51c$, we fit the data of Fig.~\ref{fig:408}(a) with a Lorentzian lineshape with the same center frequency and width as in (b). The confidence band of the fit gives a lower bound for the autoionization lifetime of 2 ms \cite{SI}.

We have prepared the circular state of strontium with $n=51$ and observed extremely long autoionization lifetimes when the ionic core electron is excited to any of the metastable $4D$ states. This is in contrast with the case of Barium \cite{roussel_observation_1990}, where the corresponding $5D_{5/2},21c$ state had only a microsecond lifetime.  The measured lifetimes (5 ms and 2 ms for the $4D_{3/2}$ and $4D_{5/2}$ states respectively) are limited by the statistical uncertainty on the survival probability. The measurement precision could be considerably improved with cold strontium atoms. 

These results open the way to optical cooling and trapping of long-lived circular state atoms with considerable impact on quantum metrology and quantum simulation.  The interaction between the core and Rydberg electron leads us to envision quantum logic operations between them. Importing optical manipulation techniques developed for ion trap experiments \cite{Leibfried_quantum_2003} makes it possible to use these quantum gate operations to detect or manipulate the outer Rydberg electron, leading to a fascinating interface between optical and mw quantum bits.

\section{Acknowledgement}

We thank M. Poirier, Ch. Koch and S. Patsch for fruitful discussions,  I. Dotsenko, C. Sayrin, P. M\'ehaignerie for experimental support. 
Financial support from the Agence Nationale de la Recherche under the project  ``SNOCAR'' (167754) is gratefully acknowledged. This publication has received funding from the European Union's Horizon 2020 under grant agreement No 786919 (Trenscrybe) and 765267 (QuSCo).

\section{Supplementary Information}
\appendix 

We provide more details about the experimental set-up, the singlet-triplet probes, the numerical model and we present the ionization signal obtained in the presence of resonant 422 nm light. 

\subsection{Detailed set-up description}

\paragraph{Rydberg excitation lasers}
The atoms are excited to the Rydberg states by three lasers at 461 nm, 767 nm and 896 nm that intersect the atomic beam in the center of the electrode structure, at a $90^\circ$ mutual angle and at a $\pm 45^\circ$ angle with the atomic beam. The 896 nm and 767 nm beams are collinear, with orthogonal linear polarizations. The linear polarization of the 896 nm laser beam is aligned with the direction of the circularly polarized 461 nm laser beam. The frequencies of the 896 nm and 767 nm lasers are both locked on an ultra-stable optical cavity. The 461 nm beam is produced by frequency doubling a 922 nm laser, frequency-locked onto the same ultra-stable cavity.

The powers of the 461 nm, 767 nm and 896 nm laser beams are 0.9 mW, 7 mW and 36 mW respectively. Their gaussian diameters (1/$e^2$ of the intensity) are 900 $\mu$m, 1100 $\mu$m and 490 $\mu$m respectively. The spatial extension of the atomic sample is defined by the $\sim$1 mm diameter of the atomic beam, by the transverse profile of the smallest-diameter 896 nm laser and by the duration of the laser excitation pulse ($1\; \mu$s, programmed at $t=0$). Due to the Doppler effect, the laser frequencies select a narrow velocity class in the broad thermal beam distribution. We measure the excited atoms velocity distribution by recording the arrival times of the Rydberg atoms at the detector (see Fig; \ref{fig:tof}). We get $\langle v \rangle = 418$ m/s and $\Delta v = 9$ m/s. 

\begin{figure}
  \centering
    \includegraphics[width=\linewidth]{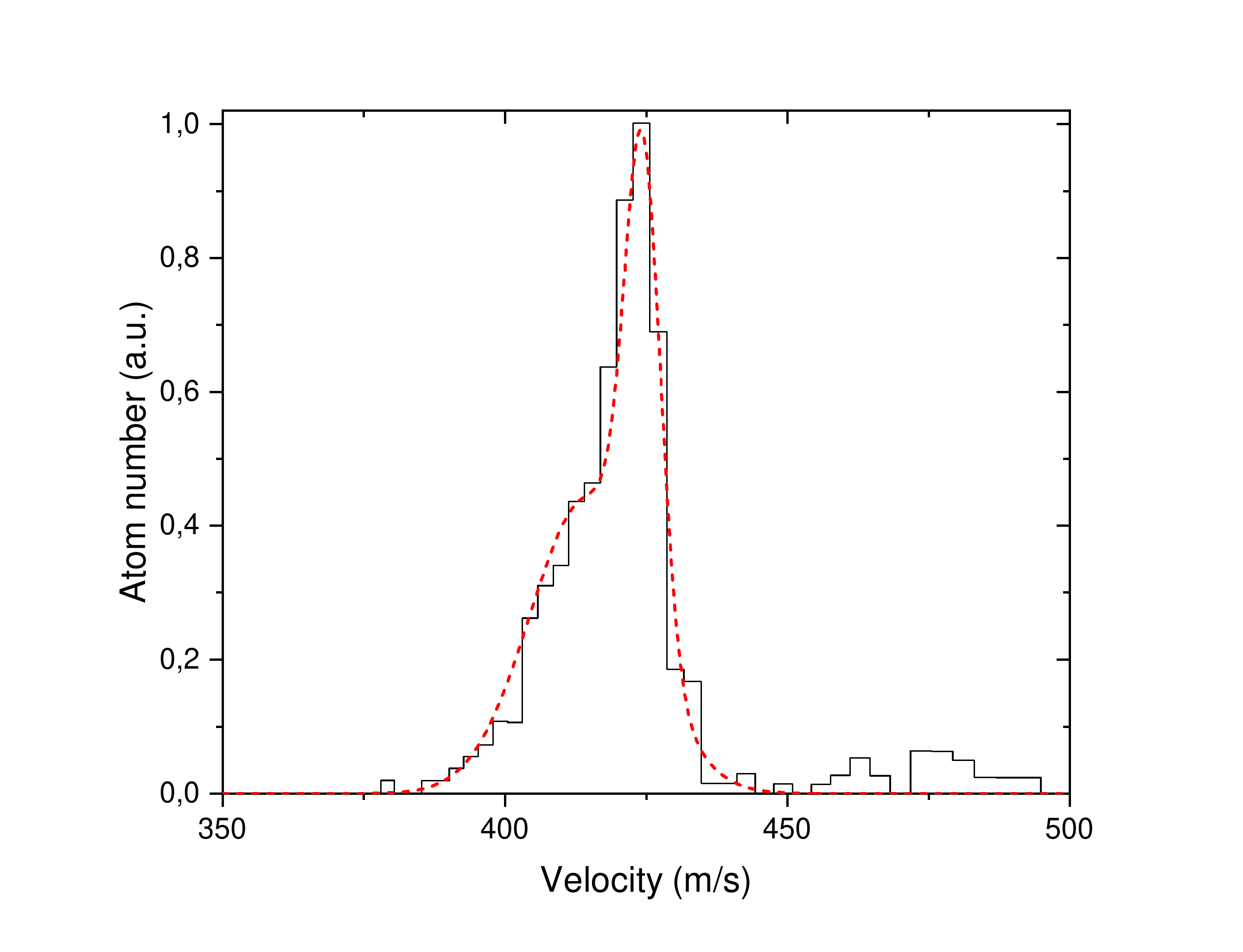}
  \caption{Velocity distribution of the atoms. The solid black line is the experimental signal deduced from the measurement of the atoms arrival times at the detector, the dotted red line corresponds the model used in the simulation.}
  \label{fig:tof}
\end{figure}

\paragraph{Electric field}

During the laser excitation and the mw preparation pulse M, an horizontal electric field is applied along the direction of the 461 nm laser. At $t=2.4\ \mu$s, we rotate adiabatically (in $0.5 \ \mu$s) the electric field to align it with the vertical axis $Oz$ and set its amplitude at $F_1=1.1$ V/cm, leading to an optimum circularization efficiency for the chosen rf frequency $\omega_\mathrm{rf}/2\pi=110$ MHz. At $t=4.2\, \mu$s, we increase the amplitude to $F= 2.1$ V/cm. This value is arbitrarily chosen large enough to ensure that the circular state is well separated in energy from the Stark level with $m=n-2$ when the atoms interact with the 422 nm laser beam. At $t=12.3\, \mu$s, we decrease the field amplitude to bring the atom back in resonance with the rf for the decircularization pulse. Finally, we increase again the field to $F=1.4$ V/cm before applying the microwave probe pulse P$_1$ or P$_2$.
\paragraph{Radiofrequency pulse}

In order to transfer the atom from the $m=2$ state to the circular state, we apply a constant amplitude radio-frequency pulse of duration $\tau$. We use the mw probe pulse P$_1$ to monitor, after the rf circularization pulse, the populations of the levels of the lowest diagonal of Stark manifold [connected with the blue dashed line in Fig. 1(a)].  Fig. \ref{fig:rabirf} presents, as a function of $\tau$, the number of atoms detected in $n=49$ when the probe mw pulse is set at resonance with the circular state, the $m=n-2$ and the $m=n-3$ states of the lower diagonal, respectively. The signals are very similar to those obtained in \cite{signoles_coherent_2017} for rubidium. We observe that the population of the circular state is optimal for $\tau= 188$ ns. 

\begin{figure}
  \centering
    \includegraphics[width=\linewidth]{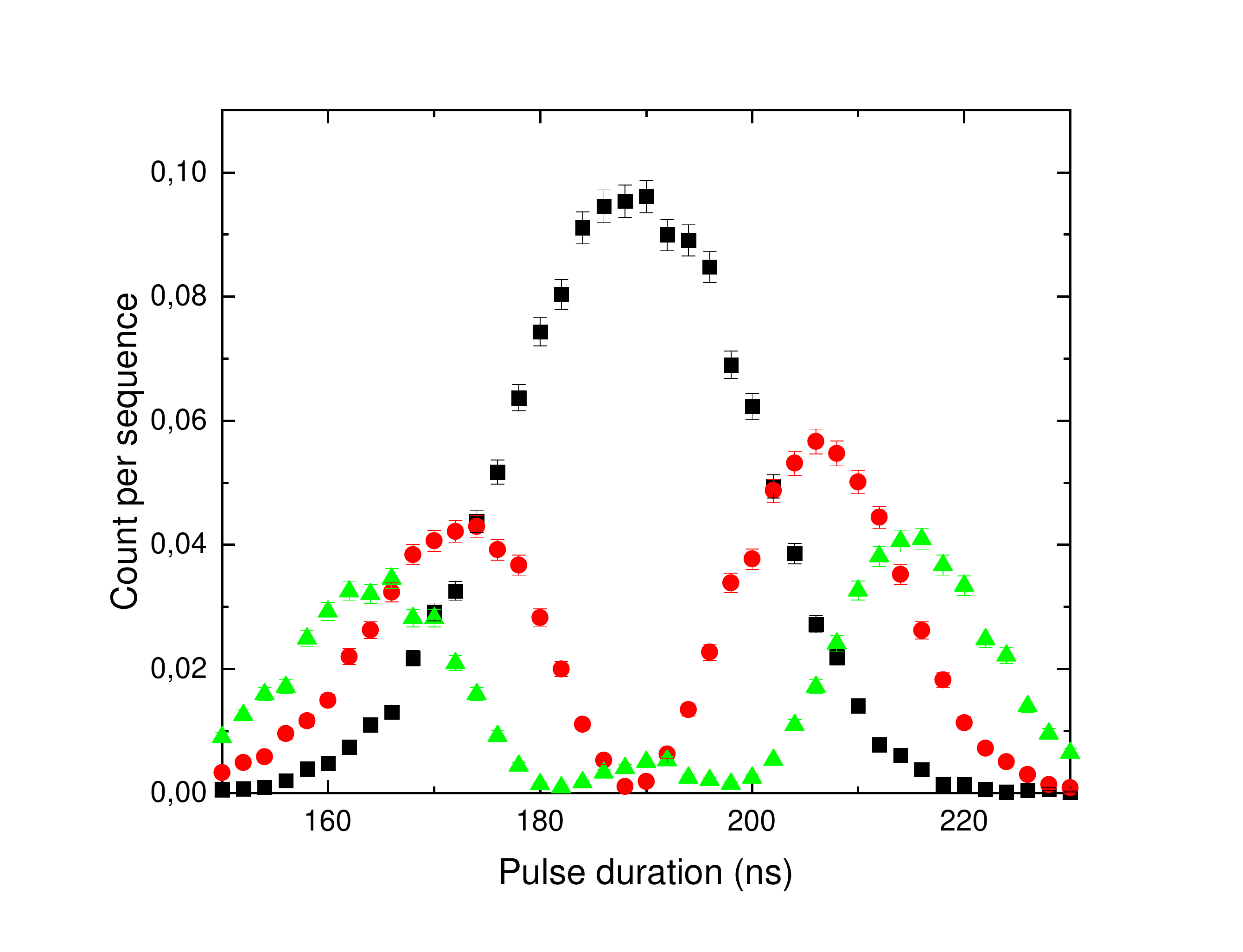}
  \caption{Number of atoms detected in the $n=49$ manifold when we apply a microwave probe pulse that selectively transfers the $n=51$ circular state (black), the lowest $m=49$ energy level of the $n=51$ manifold (red) and the lowest $m=48$ energy level of the $n=51$ manifold (green) into the $n=49$ manifold, as a function of the duration $\tau$ of the circularization rf pulse. We clearly observe a maximum of the population of the $51c$ level for $\tau=188$~ns. There, the residual population of the m=48 level is only  about  $5\%$.}
  \label{fig:rabirf}
\end{figure}

\paragraph{Ionic core excitation lasers}

The 422 nm laser beam is produced by a laser diode. It is locked to a moderate-finesse transfer cavity, which is itself referenced to the frequency of the 767 nm laser, shifted by an electro-optic modulator (EOM). We thus scan the frequency of the 422 nm laser by scanning the source driving the EOM. From the transmission signal of the transfer cavity, we estimate the linewidth of the 422 nm laser to be 3 MHz. It is small as compared to the linewidth of the $5S_{1/2} \rightarrow 5P_{1/2}$ transition of Sr$^+$ ($\sim20$ MHz). We also observe a slow residual drift of the laser frequency ($\sim 15$ MHz over a few hours of data acquisition) due to drifts in the locking system. To eliminate this drift, we record before and after each curve of Fig. \ref{fig:spectro} a reference curve, in which we always measure the transfer into the $D_{3/2}$ state as a function of the laser frequency $\nu_\mathrm{l}$ in the same conditions. We measure the resonance frequency $\nu_\mathrm{l}^0$ in the two reference spectra and plot the signals in Fig. \ref{fig:spectro} as a function of $$\Delta\nu_\mathrm{l}= \nu_\mathrm{l} - \frac 1 2 (\nu_\mathrm{l,after}^{0}+\nu_\mathrm{l,before}^{0}).$$
The residual uncertainty on the laser frequency due to the drift \emph{during} the scan can be estimated from the difference $|\nu_\mathrm{l,after}^{0}-\nu_\mathrm{l,before}^{0}|$. It is typically 5 MHz.

The 422 nm beam is sent parallel to the 461 nm beam. The two beams are separated by 2 mm, so that the atoms cross the 422 nm laser 6 $\mu$s after being excited to the Rydberg state. The 422 nm beam has a diameter of 850 $\mu$m at the position of the atomic beam. This laser is pulsed (duration $5\ \mu$s). This duration being larger than the transit time of the atoms across the beam, the latter sets the interaction time between the atoms and the 422 nm laser.

The 1092 nm repumper light is produced by a laser diode locked to the ultra-stable reference cavity. The output of the laser diode is split into two beams of orthogonal linear polarizations. They are frequency-shifted by two independent acousto-optic modulators (AOM) driven at frequencies differing by 3 MHz. The two beams are  recombined on a polarizing beam splitter before being sent onto the atoms. As a result, the final polarization of the laser is modulated at 3 MHz. This makes it possible to efficiently repump all possible dark states in the $D_{3/2}$ level \cite{berkeland_Destabilization_2002}. The 1092 nm beam has a diameter of 1500 $\mu$m at the position of the atom, and a power of about 10 mW for each polarization.  

The 408 nm laser beam is produced by a frequency-doubled free-running Ti:Sa laser sent along the path of the 422 nm beam. The 408 nm light propagates in a fiber without polarization control and is combined with the 461 nm beam  on a polarizing beam splitter. The random rotations of the 408 nm light polarization  thus lead to fluctuations of the laser power, limiting the precision with which we control the power $P^\prime$ sent to the atoms. We measure typically $P^\prime\sim 1.5$ mW, for a beam diameter of 850 $\mu$m at the atomic beam position. The 408 nm laser is continuous wave, but  its position along the atomic beam is set downstream enough from the Rydberg excitation region to make sure that the atoms do not interact with this light before they are already transferred into the circular state. Experimentally, we first perform the experiment where we record the number of atoms $N$ and $N^c$ in the presence of 408 nm laser as a function of its frequency. We then block the 408 nm beam to measure the numbers of atoms $N_0$ and $N^c_0$ in the absence of blue light.

\subsection{Singlet-triplet probes}

\begin{figure}
  \centering
    \includegraphics[width=\linewidth]{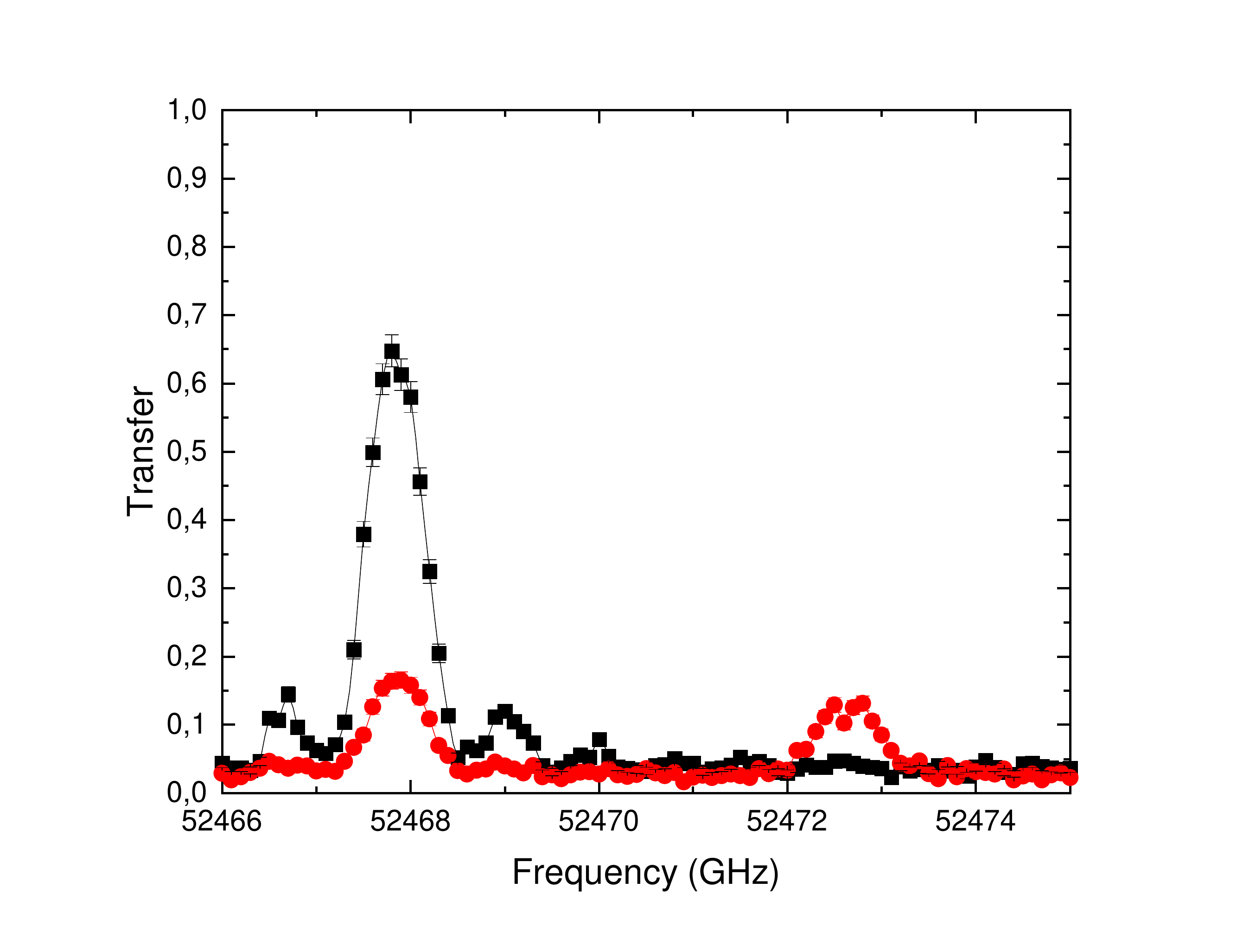}
  \caption{Number of atoms transferred into in the $n=49$ manifold by a microwave pulse applied after the decircularization rf pulse, normalized by the number of atoms prepared in the $51m2$ state, as a function of the microwave frequency. The black dots correspond to the experimental signal without the 422 nm laser, the red dots show the signal when we apply the laser on resonance. The 1092 nm light is on in both cases. The error bars are statistical. The solid lines are a guide to the eye.}
  \label{fig:stsi}
\end{figure}

In order to determine the frequency of the microwave probes P$_2$ used to discriminate between the triplet and singlet states, we first record the microwave spectrum near the transition frequency between the lowest $m=2$ state of the $n=51$ manifold (``$51m2$") and the lowest $m=1$ state of the $n=49$ manifold (``$49m1$"). Fig. \ref{fig:stsi}
 presents the number of atoms detected in $n=49$, as a function of the mw frequency, $\nu_{\mathrm{mw}}$ with or without applying the 422 nm laser onto the atoms, in the presence of repumper light. This laser transfers singlets to triplets by optical pumping as seen in the main text. When no 422 nm laser is applied, we observe a single resonance at $\nu_{\mathrm{mw}} = 52.4678$ GHz. It corresponds to the transition from the $51m2$ to the $49m1$ singlet states. When the laser is applied, we observe an additional resonance attributed to the transition between the $51m2$ and $49m1$ triplet states. Note that the mw pulse duration, $0.6\  \mu$s, is too short to resolve the internal structure of these triplet states. We use microwave probe pulses at these two frequencies to measure the populations of the $5S_{1/2},51c$ singlet and triplet states after the atoms have interacted with the 422 nm laser. 

The proportionality factors between $5S_{1/2},51c$ singlet and triplet populations and number of atoms detected after the singlet or triplet probe depends on the transfer efficiencies of the decircularization rf pulse and of that of the mw probes. In the case of the singlet probe, this factor is easily measured, since we know that, without 422 nm laser, the atoms are all in a singlet state. We thus simply normalize the atomic count by the number of atoms detected with that probe when we do not apply the 422 nm laser. In the case of the triplet probe, we calibrate the proportionality factor by assuming that, for a large 422 nm laser power and in the presence of repumper light, the population of the triplet state is 75~\%.

\subsection{Numerical model}

Our numerical model is based on the integration of the density matrix evolution under the action of the 422 nm laser beam detuned by $\Delta/2\pi$ from the $5S_{1/2},51c \rightarrow 5P_{1/2}, 51c$ transition frequency. We take into account the states 
\begin{eqnarray*}
\ket{5S_{1/2},m_J=-1/2}\ket{ 51c,\downarrow} & \  ,\ &   \ket{5P_{1/2},m_J=-1/2}\ket{ 51c, \downarrow}\   ,\\
\ket{5S_{1/2},m_J=-1/2}\ket{ 51c, \uparrow} & \   ,\ &  \ket{5P_{1/2},m_J=-1/2}\ket{  51c,\uparrow} \   , \\
\ket{5S_{1/2},m_J=+1/2}\ket{  51c,\downarrow} & \   ,\ &  \ket{5P_{1/2},m_J=+1/2}\ket{  51c,\downarrow}  \     ,\\
\ket{5S_{1/2},m_J=+1/2}\ket{  51c,\uparrow} & \   ,\ &  \ket{5P_{1/2}m_J=+1/2}\ket{  51c,\uparrow} \   \ , \\
\end{eqnarray*}
where the spin quantization axis is chosen along the direction of the 422 nm laser. We also include a ninth state ``$\ket{ D_{3/2}}$'', which formally represents all the $D_{3/2}$ levels in which the $5P_{1/2}$ sub-levels can decay with the same rate. Fig. \ref{fig:model} represents the couplings and decay channels  included in the model. Note that the spin of the 51c electron is not coupled to the 422 nm light and is spectator during excitation and fluoresence. The decay rates from  $\ket{5P_{1/2},m_J }$  to $\ket{5S_{1/2},m_J }$ and $\ket{5S_{1/2},-m_J }$ are $\Gamma_\pi$ and $\Gamma_\sigma$  respectively ($\Gamma_\sigma /2 \pi= 2  \Gamma_\pi /2 \pi= 13.6 $ MHz), $\gamma$ is the decay rate of $\ket{5P_{1/2},m_J}$ into $\ket{ D_{3/2}}$ ($\gamma/2 \pi = 1.2$ MHz). Finally,   $\Omega_+$ and $\Omega_-$ are the Rabi frequencies associated to the $5S_{1/2},51c \rightarrow 5P_{1/2}, 51c$ transitions with $\Delta m_j = 1$ or $-1$ respectively. It is proportional to the amplitude of the $\sigma_+$ or $\sigma_-$ component of the 422 nm laser at the position of the atoms, and thus to the square root of the local laser intensity $I(x,z)$. According to the measured laser polarization, we set $\Omega_+ = \Omega_-/6$.

\begin{figure}
  \centering
    \includegraphics[width=\linewidth]{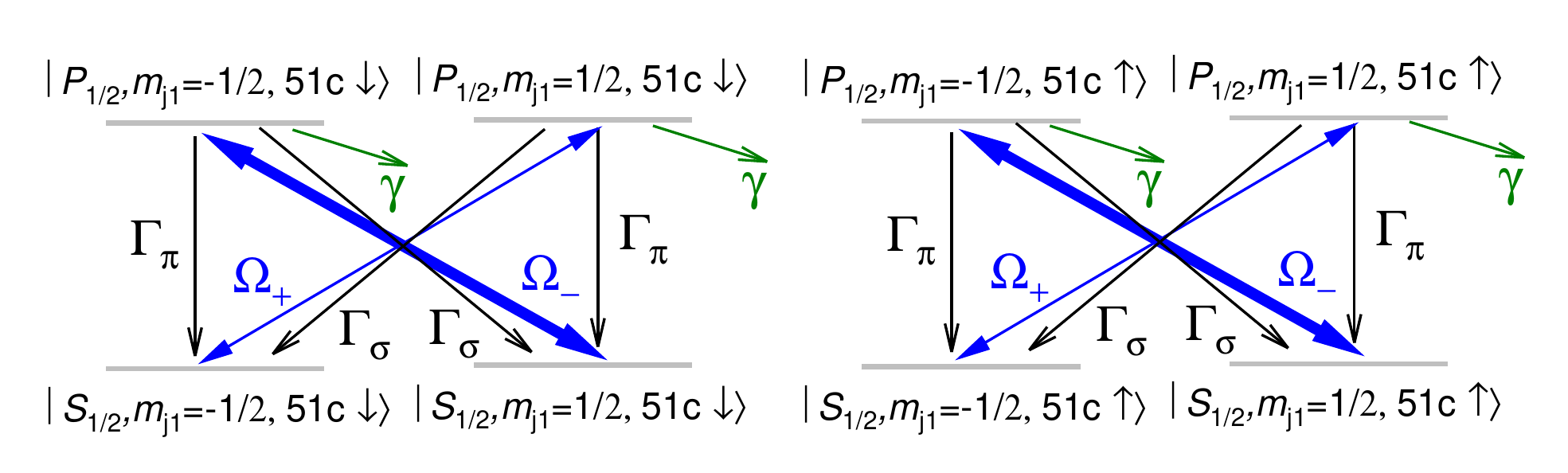}
  \caption{Scheme of the levels, couplings and decay channels included in the numerical model. The green arrows $\gamma$ represent the decay of the $P_{1/2}$ sublevels into the $D_{3/2}$ states.}
  \label{fig:model}
\end{figure}

In the experiment, the atoms move across the gaussian profile of the 422 nm laser beam at a velocity $v_0 = 418\pm 9$ m/s along the $(Ox)$ direction. The Rabi frequencies $\Omega_+$ and $\Omega_-$ thus vary accordingly along the atomic trajectory. We have numerically checked that the dynamics of the atom crossing a beam of effective gaussian diameter $\sqrt 2 w_x = 580 \, \mu$m (the factor $\sqrt 2$ takes into account the angle between the atomic beam and the laser) can be approximated by that of an atom staying at the point of maximum intensity, $I(0,z)$, for an effective time $t_\mathrm{eff}= 1.8\,\mu$s . 

We simulate the experimental data, by first computing the probabilities $p_S(I(0,z),\Delta)$ and $p_T(I(0,z),\Delta)$ for an atom initially in the singlet state $$\frac 1 {\sqrt 2} \left [\ket{5S_{1/2},m_J=-\frac 1 2}\ket{ 51c \uparrow}-\ket{5S_{1/2},m_J=\frac 1 2}\ket{  51c\downarrow} \right ]$$ to end up in a singlet or triplet state respectively, after interacting for at time $t_\mathrm{eff}$ for a laser of intensity $I(0,z)$ and detuning $\Delta$. (Note that the interaction is followed by an integration without lasers for $t\gg \gamma^{-1}$, so that the population of the $5P_{1/2}$ state has fully decayed into the $5S_{1/2}$ and $5S_{1/2}$ levels).

We then average the result over the spatial extension of the atomic sample along the $Oz$ axis to take into account the profile of the laser beam in the direction perpendicular to the atomic beam. We convolute the result with the velocity distribution of the atoms (modeled by the red curve on Fig.~\ref{fig:tof}), thus taking into account the Doppler effect on the laser detuning $\Delta$. We finally convolute the frequency response with a 3 MHz standard deviation gaussian to account for the frequency width of the laser. 
We finally get the average probabilities $\bar p_S(I_0,\bar\Delta)$ and $\bar p_T(I_0,\bar\Delta)$ for the atom to be in a singlet or triplet state before the decircularization pulse as function of the intensity $I_0$ at the center of the laser beam and  of the detuning $\bar\Delta$ between the frequency of the $5S_{1/2},51c \rightarrow 5P_{1/2}, 51c$ transition and that of the laser field experienced by an atom at velocity $v_0$.

Fig. \ref{fig:spectroprobe} compares the experimental data to the computed values of $\bar p_S(I_0,\bar\Delta)$ and $\bar p_T(I_0,\bar\Delta)$ (assuming $\Delta\nu_\mathrm{l}=\bar\Delta$). The only free parameter  is the proportionality factor between the peak intensity $I_0$ and the power $P$ of the laser beam, that we fit to be 26 \% smaller than the result of an {\sl ab initio} calculation from the estimated size of the laser beam at the atomic position. 

\begin{figure}
  \centering
    \includegraphics[width=\linewidth]{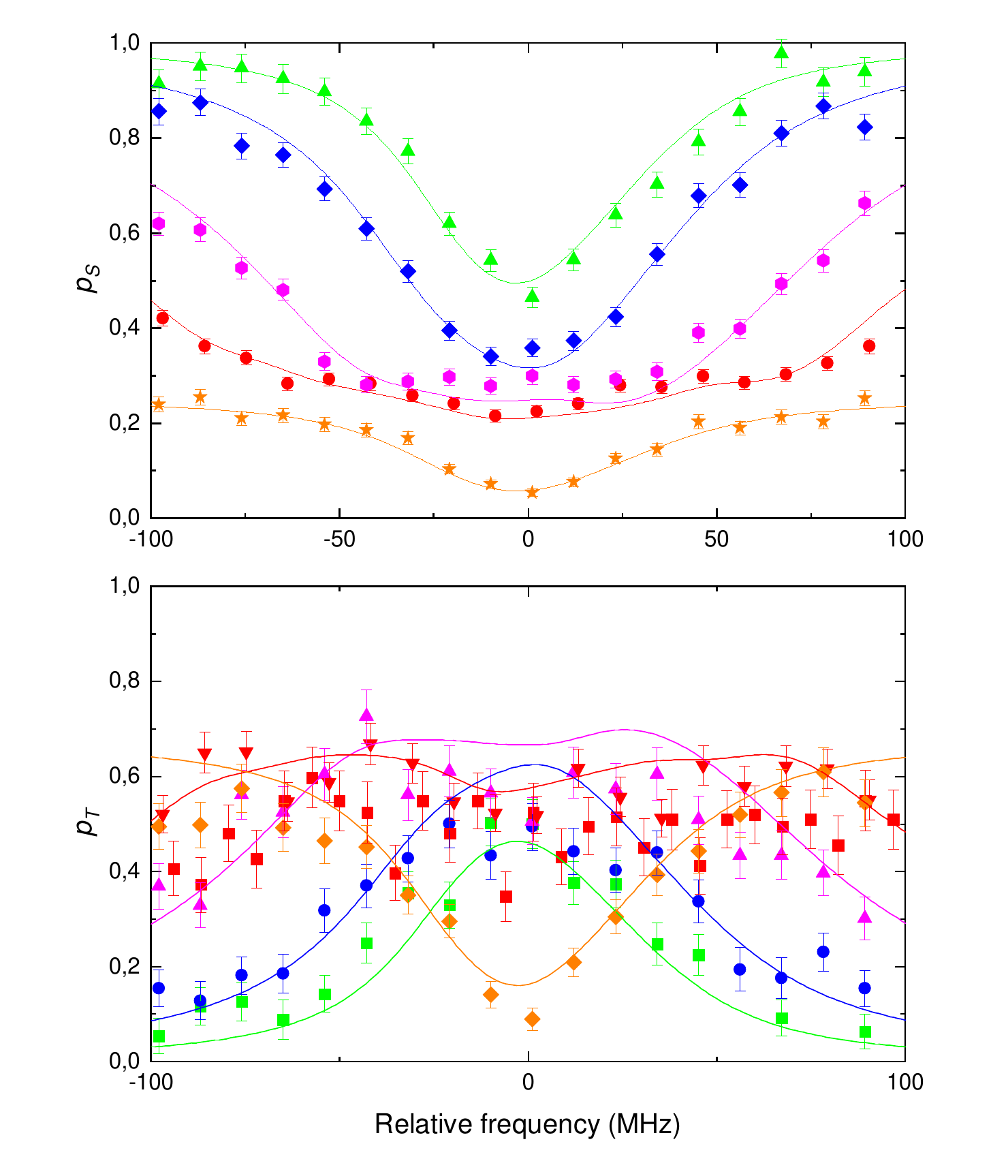}
  \caption{Probability $p_S$ (upper frame) and $p_T$ (lower frame) for the atom to be in the $5S_{1/2},51c$ singlet and triplet state after the 422 nm laser pulse as function of the laser frequency for different powers $P$ (green : $9.6\, \mu$W, blue : $19\, \mu$W, pink: $43\, \mu$W, red : 0.29 mW, orange : 4 mW). The points are experimental data (with statistical error bars), the lines present the results of the numerical model.}
  \label{fig:spectroprobe}
\end{figure}

\subsection{Ionization signal}

Fig. \ref{fig:dig} compares the ionization signals obtained with and without applying the 422 nm light, when the laser frequency is tuned to resonance ($\Delta\nu_\mathrm{l}=0$). Fig.  \ref{fig:dig}(a) presents the signal obtained without the decircularization rf pulse. We observe that the ionization peak of the circular state is impervious to the laser pulse. In contrast, the right-hand side peak , corresponding to residual $5s52f\ ^1F_3\ m=2$ atoms, disappears when the 422 nm laser is switched-on. The ionization signal does not depends on presence of repumper light, demonstrating that the state $4D_{3/2},51c$ do not experiences sizable autoionization. Fig. \ref{fig:dig}(b) presents the signal when we apply the two rf pulses. Although the second rf pulse brings back the circular state into low-$m$ states, it does not succeed in transferring all the atoms into $m=2$. Some population is left in intermediate $m$ levels (small peak in the center of the curve). When the 422 nm pulse is applied, the ionization peak on the right side (low-$m$ states) almost completely disappears and only the center peak (intermediate $m$s) remains. This is consistent with a loss due to autoionization, as it is expected to be less efficient for Rydberg states with higher $m$s. In the presence of repumper light, the ionization peak on the right reappears, showing that the atom loss is due to population transfer in the $4D_{3/2}, 51c$ state. 

\paragraph{Background $5s52f\ ^1F_3\ m=2$ suppression} The ionization threshold of the $5s52f\ ^1F_3\ m=2$ level is close to that of the $51, m=2$. The remaining $5s52f\ ^1F_3\ m=2$ atoms thus contribute to the number of atoms that we detect at the threshold of the $51, m=2$. In order to substract this background, we run two different sequences. We first measure the $5s52f\ ^1F_3\ m=2$ background by counting the number of atoms detected at the $m=2$ threshold without the decircularization rf pulse. We then measure the number of atoms detected at the $m=2$ threshold when we apply the decircularization rf pulse, which includes both the $n=51$ Rydberg atoms and the background from  $5s52f\ ^1F_3\ m=2$. As the $5s52f\ ^1F_3\ m=2$ atoms are impervious to the rf pulses these two measurements can be used for background substraction.}

\begin{figure}
  \centering
    \includegraphics[width=\linewidth]{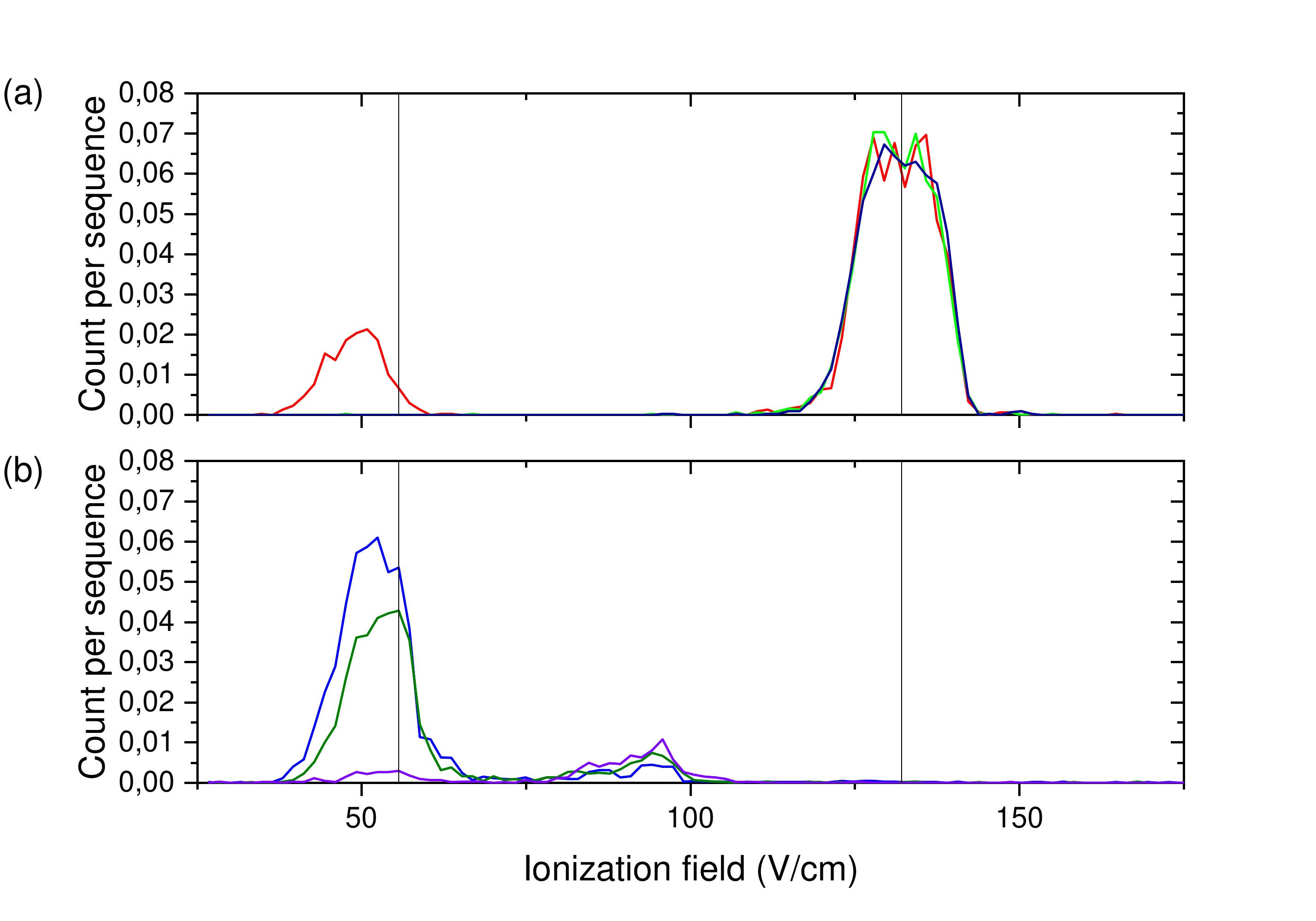}
  \caption{(a) Ionization signals. Number of detected atoms as a function of the ionization voltage when we prepare the circular state (red), when we additionally apply a resonant 422 nm laser pulse (green), and when we apply a pulse of resonant 422 nm laser in the presence of 1092 nm repumper light (black). (b) Ionization signal obtained when we apply both circularization and decircularization pulses without the 422 nm laser (blue), with the 422 nm laser and without the 1092 nm light (purple) and with both the 422 nm and the 1092 nm lights (olive). The vertical black lines indicate the ionization thresholds used to detect the $51,m=2$ (left) and 51 circular states (right). The threshold for the $51, m=2$ state is not centered on the ionization peak of $51, m=2$ in order to limit the background of $5s52f\ ^1F_3\ m=2$ atoms detected as $51, m=2$ [left peak on the red signal of (a)].}
  \label{fig:dig}
\end{figure}

\subsection{Autoionization lifetime estimation}

\paragraph{Ionic core level $4D_{3/2}$} 
We estimate the autoionization lifetime of the $4D_{3/2},51c$ level from the data of Fig. \ref{fig:spectro}. The survival probability $N^c/N^c_0$ is given by 
$$\frac{N^c}{N^c_0} = P_{S_{1/2}} + P_{D_{3/2}}\exp\left(-\frac{t}{\tau_{D_{3/2}}}\right)$$
where $P_{S_{1/2}}$ is the probability to be in the ${S_{1/2}}$ state, $P_{D_{3/2}}$ is the probability to be in the ${D_{3/2}}$ state, $\tau_{D_{3/2}}$ the autoionization lifetime of the ${D_{3/2}}$ state, and $t= 140\, \mu$s is the time during which the atom is in the ${D_{3/2}}$ state (from the time of the 422 nm laser pulse to the time when the atom reaches the detector). Since $P_{S_{1/2}} = 1-  P_{D_{3/2}}$, we get 
$$\frac{N^c}{N^c_0} = 1- P_{D_{3/2}}\left[1-\exp\left(-\frac{t}{\tau_{D_{3/2}}}\right)\right]$$
We thus fit the data of Fig. \ref{fig:spectro}(a) with the formula 
$$\frac{N^c}{N^c_0} = 1- \beta P_{D_{3/2}}(\Delta\nu_\mathrm{l})$$
where $P_{D_{3/2}}(\Delta\nu_\mathrm{l})$ results from the numerical model and $\beta$ is the fit parameter. From the fit 95 \% confidence band, we find that 
$$0.004 < \beta < 0.028$$ 
providing a lower bound on the autoionization lifetime $\tau_{D_{3/2}} > 5$ ms.

\paragraph{Ionic core level $4D_{5/2}$} 

We estimate the autoionization lifetime of the $4D_{3/2},51c$ level from the data of Fig. \ref{fig:408}. In the case of the 408 nm core excitation, the atom can decay from the $P_{3/2}$ into the $D_{3/2}$ and $D_{5/2}$ states.
We cannot here rely on a numerical model to determine the population of the $D$ states. We estimate it from the dip in the $N/N_0$ signal in Fig. \ref{fig:408}(b). 
At the lowest point of the Lorentzian fit, the ratio $N/N_0$ is 8 \%. Part of the atom count reduction can be due to a transfer of the atoms into the $5S_{1/2},51c$ triplet state, which are detected with a 7 \% smaller probability. Nevertheless, it means that at least 85 \% of the population is optically pumped in the $D$ states. From the branching ratio $r/r^\prime$ between the decay channels from $P_{3/2}$, we get that at least 76 \% of the population is in the $D_{5/2}$ level and 9 \% is in $D_{3/2}$. 

The survival probability of the circular state is then given by
$$\frac{N^c}{N^c_0} = P_{S_{1/2}} + P_{D_{3/2}}e^{-t/\tau_{D_{3/2}}}+ P_{D_{5/2}}e^{-t/\tau_{D_{5/2}}}$$
To get a lower bound on $\tau_{D_{5/2}}$, we assume that $\tau_{D_{3/2}} = \infty$ and thus that all the loss is due to the population in the ${D_{5/2}}$ state. 
From the Lorentz fit of Fig. \ref{fig:408}(a), we get that, at the minimum of the curve  $ {N^c}/{N^c_0}  > 0.946$ with more than 95 \% probability. We then deduce $\tau_{D_{5/2}}>2$ ms.

\end{document}